\documentclass[12pt]{article}
\makeatletter

\@addtoreset{equation}{section}
\makeatother

\begin{document}

\pagestyle{plain}
\setcounter{page}{1}
\begin{titlepage}

\leftline{\tt hep-th/0203047}

\vskip -.8cm

\rightline{\small{\tt TU-647}}

\begin{center}

\vskip 2 cm

{\LARGE\bfseries Eight Dimensional Noncommutative Instantons 
and D0-D8 Bound States with $B$-field}

\vskip 2cm
{\large Yoshiki Hiraoka}

\vskip 1.2cm

{\it Department of Physics, Tohoku University}

{\it Sendai 980-8578, JAPAN}

\vskip .7cm
{\tt hiraoka@tuhep.phys.tohoku.ac.jp}

\vskip 1.5cm

{\bf Abstract}
\end{center}

\noindent

We construct some classes of instanton solutions of eight dimensional
 noncommutative ADHM equations generalizing the solutions of eight dimensional
 commutative ADHM equations found by Papadopoulos and Teschendorff,
 and interpret them as supersymmetric $D0$-$D8$ bound states
 in a NS $B$-field.
Especially, we consider the $D0$-$D8$ system with anti-self-dual 
$B$-field preserving 3/16 of supercharges. 
This system and self-duality conditions are related with the group $Sp(2)$
 which is a subgroup of the eight dimensional rotation group $SO(8)$.

\end{titlepage}

\newpage


\section{Introduction}

Recently noncommutative geometry has appeared in the context of 
M(atrix)-theory compactifications~\cite{cds} and 
the $D$-branes in a constant NS $B$-field~\cite{sw}.
In particular, the $D$-branes in a constant NS $B$-field have 
attracted much interest in the understanding of non-perturbative 
aspects of string theory.
The noncommutative Yang-Mills theory which appears as an 
effective world-volume field theory on the $D$-branes with a $B$-field
has an interesting feature that the singularity of the instanton moduli space 
is naturally resolved~\cite{ns}. 

Four dimensional $U(N)$ $k$-instanton is realized as $k$ $D0$-branes 
within $N$ $D4$-branes in type IIA string theory.
When we turn on an anti-self-dual constant $B$-field, 
for preserving 1/4 of supercharges,  
the instanton moduli space is resolved, and $D0$-branes 
cannot escape from $D4$-branes.
From the viewpoint of $D0$-brane theory, 
the moduli space of vacua of the Higgs branch coincides with the moduli space 
of instantons and the anti-self-dual $B$-field corresponds to  
the Fayet-Iliopoulos (FI) parameters.
If the FI parameters are non-zero, the $D0$-$D4$ system can not enter the 
Coulomb branch through the small instanton singularity.

On the other hand, it is also of interest to generalize the above case 
to higher dimensional systems in the context of both brane dynamics and the 
brane world-volume theories.
These systems with a constant $B$-field are considered 
from various points of view in~\cite{cimm, park, witten, fio, ohta}, 
and are shown equivalent by T-duality to the systems 
that are rotated branes at angles by several 
authors~\cite{ohtan, ohta, pt, pt2, ggpt}.
In particular they showed that there are three cases  
preserving respectively 1/16, 1/8 and 3/16 of supercharges 
in the $D0$-$D8$ systems with a $B$-field that must satisfy certain 
relations in each case. 
Concretely in the $D0$-$D8$ system, the constant $B$-field satisfies the 
extended ``self-dual'' conditions given in~\cite{cdfn, ward} 
which associate with the subgroup $Spin(7), SU(4)$ and $Sp(2)$ 
of the eight dimensional rotational group $SO(8)$.
These $B$-fields preserve 1/16, 1/8 and 3/16 of supercharges respectively.

The ADHM construction is a powerful tool to construct self-dual Yang-Mills 
instantons~\cite{adhm, cg}.
Especially in four dimensions we know that the instanton moduli space 
and the ADHM moduli space completely coincide. 
In eight dimensions, the ADHM construction~\cite{cgk} 
of ``self-dual'' instantons associated with the group $Sp(2)$ 
is known, 
but we does not know whether this ADHM construction gives all solutions 
of ``self-dual'' equations associated with the group $Sp(2)$ .
However, some simple solutions of eight dimensional ADHM equations are already 
known in~\cite{cgk, pt}.
Eight dimensional noncommutative ADHM equations were proposed in~\cite{ohta}, 
but these equations are difficult to solve even in the simple case. 
Some properties of moduli space of these equations 
were also discussed in~\cite{hio}.

This paper is organized as follows. 
In section 2, we describe eight dimensional noncommutative 
Yang-Mills instantons associated with the group $Sp(2)$ 
as $D0$-$D8$ bound states in a $B$-field
using the noncommutative eight dimensional ADHM construction. 
In this case, the $B$-field is also necessary to satisfy the extended 
``(anti-)self-dual'' relations.
We can construct some new classes of non-trivial solutions 
extending the eight dimensional ADHM constructions of~\cite{cgk} 
and the solutions found by Papadopoulos and Teschendorff~\cite{pt} 
in the commutative case, 
and interpret them as $D0$-$D8$ bound states.
The final section is devoted to discussions.


\section{Noncommutative instantons on $\mathbf{R}^8$ 
as $D0$-$D8$ bound states with a $B$-field}

In this section, we construct noncommutative instantons on $\mathbf{R}^8$ 
 and interpret them as supersymmetric $D0$-$D8$ bound states
 with a certain  $B$-field.
This noncommutativity is induced by a constant NS $B$-field 
on the $D8$-brane.
In the following, we consider the case of the gauge group $U(N)$ with 
the instanton number $k$ since these states correspond to the 
bound states of $k$ $D0$-branes and $N$ $D8$-branes.

We also construct some solutions of the eight dimensional 
noncommutative ADHM equations.
Eight dimensional noncommutative ADHM equations were proposed in~\cite{ohta}, 
but it is difficult to solve those equations even in the $U(2)$ case.
So we try to solve the equations 
by generalizing the solutions found by Papadopoulos 
and Teschendorff~\cite{pt, pt2} in the commutative case.


\subsection{Eight dimensional ADHM construction}

In this subsection, we consider the extended ADHM construction of the 
eight dimensional ``self-dual'' instantons associated with the $Sp(2)$ 
group given in~\cite{ward, cgk}.
This construction of instantons is the slight extension of 
the four dimensional ADHM construction.
When we take $\mathbf{B}=0\,,$ or $\mathbf{B}'=0$ which are defined 
in the following, 
we will reproduce the four dimensional ADHM equations. 
Some simple solutions to these equations
 were constructed in~\cite{ns, nekrasov, 
furuuchi, furuuchi2, furuuchi3} and others.

In order to treat the eight dimensional space, it is useful to regard 
eight coordinates of $\mathbf{R}^8$ as two  
quaternionic coordinates 
\begin{equation}
\mathbf{x}=\sum_{\mu =1}^8 \tilde{\sigma}_{\mu}x^{\mu}=
\left( \begin{array}{@{\,}cc@{\,}}
   z_2 & z_1\\
  -\bar{z}_1  &  \bar{z}_2
  \end{array}  \right)\,,\quad 
\mathbf{x}'=\sum_{\mu =1}^8 \tilde{\sigma}'_{\mu}x^{\mu}=
\left( \begin{array}{@{\,}cc@{\,}}
   z_4 & z_3\\
  -\bar{z}_3  &  \bar{z}_4
  \end{array}  \right)\,,
\end{equation}
where we defined the eight vector matrices
\begin{eqnarray}
\tilde{\sigma}_{\mu}&=&(\,i\,\tau_1,\,0,\,i\,\tau_2,\,0,\,i\,\tau_3,\,0,\,\textbf{1}_2,\,0\,)\,,\\
\tilde{\sigma}'_{\mu}&=&(\,0,\,\,i\,\tau_1,\,0,\,i\,\tau_2,\,0,\,i\,\tau_3,\,0,\,\textbf{1}_2\,)\,,
\end{eqnarray}
and the four complex coordinates
\begin{equation}
 z_1=x^3+ix^1\,,\quad z_2=x^7+ix^5\,,\quad 
z_3=x^4+ix^2\,,\quad z_4=x^8+ix^6\,.
\end{equation}
Using the $(N+2k)\times 2k$ matrices $\mathbf{A}\,,\,\mathbf{B}$ 
and $\mathbf{B}'$, we next define 
the Dirac-like operator 
\begin{equation}
D_z = \textbf{A}+\overrightarrow{\mathbf{B}}\cdot \overrightarrow
{\mathbf{X}}
\end{equation}
where  $\overrightarrow{\mathbf{B}}=(\mathbf{B},\mathbf{B}')$ and  
$\overrightarrow{\mathbf{X}}=(\mathbf{x},\mathbf{x}')$\,. 

If we solve the following Dirac-like equations
\begin{equation} 
D_z^{\dagger}\psi=0\,,
\end{equation} 
for the $(N+2k)\times N$ matrix $\psi$ which is normalized as 
$\psi^{\dagger}\psi =\mathbf{1}_{N\times N}$, 
we can construct the $U(N)$ gauge field as
\begin{equation}
 A_{\mu}=\psi^{\dagger}\partial_{\mu}\psi\,.
\end{equation}
Then using the relations
\begin{equation}
\Sigma_{\mu}\equiv \partial_{\mu}\overrightarrow{\mathbf{X}}=
\left( \begin{array}{@{\,}c@{\,}}
   \tilde{\sigma}_{\mu} \\
  \tilde{\sigma}'_{\mu}
  \end{array}  \right)\,,
\end{equation}
and the completeness equation
\begin{equation}
\textbf{1}_{N+2k}=\psi\psi^{\dagger}
+D_z\frac{1}{ D^{\dagger}_zD_z}D^{\dagger}_z\,,
\end{equation}
we can obtain the ``self-dual'' gauge field strength as
\begin{eqnarray}
F_{\mu\nu}&=& 2\psi^{\dagger}\left( \partial_{\left[ \mu\right. } D_z
\frac{1}{ D^{\dagger}_zD_z}  \partial_{\left. \nu \right]  }D^{\dagger}_z
\right)\psi\nonumber\\
&=& 2\psi^{\dagger}\overrightarrow{\mathbf{B}}\,
\overline{N}_{\mu\nu}\frac{1}{ D^{\dagger}_zD_z}
\overrightarrow{\mathbf{B}}^{\dagger}\psi\,,
\end{eqnarray}
where $\overline{N}_{\mu\nu}=\frac{1}{2}(\Sigma_{\mu}\Sigma_{\nu}^{\dagger}-
\Sigma_{\nu}\Sigma_{\mu}^{\dagger})$ is a ``self-dual'' tensor satisfying
\begin{equation}
  \frac{1}{2}T_{\mu\nu\rho\sigma}\overline{N}_{\rho\sigma}=
\overline{N}_{\mu\nu}\,.
\end{equation}

Here we must require that  $D^{\dagger}_zD_z$ commutes 
with $\Sigma_{\mu}$.
This is a necessary condition to obtain the ``self-dual'' gauge field strength 
on the $\mathbf{R}^8$. This condition corresponds to the eight dimensional 
ADHM equations both for commutative and noncommutative case.


\subsection{Eight dimensional noncommutative ADHM construction}

We can obtain supersymmetric $D0$-$D8$ bound states 
with a $B$-field by T-duality from intersecting $D4$-branes at four angles.
Supersymmetry condition reduces to three cases 
preserving 1/16, 1/8 and 3/16 of supercharges respectively.
In the following, we concentrate on the 3/16 BPS states. 
In this case, the $B$-field is necessary to satisfy 
the generalized ``(anti-)self-dual'' 
equations~\cite{cdfn, ward, hull} which 
are written by using the $Sp(2)\subset SO(8)$ invariant tensor 
$T_{\mu\nu\rho\sigma}$ as 
\begin{equation}
\frac{1}{2}T_{\mu\nu\rho\sigma} B^{\rho\sigma}=\lambda B_{\mu\nu}\,.
\end{equation}

As in the four dimensional case, 
it is also easy to extend the eight dimensional ADHM construction 
to noncommutative space because of its algebraic nature.
Since we define instantons as ``self-dual'' configurations, 
the ``anti-self-dual'' $B$-field is of interest from the viewpoint of 
the instanton moduli space resolution.
In this case the coordinates of $\mathbf{R}^8$ become noncommutative as
\begin{equation}
\left[ z_1,\bar{z}_1\right]=-\left[ z_2,\bar{z}_2\right]=
\left[ z_3,\bar{z}_3\right]=-\left[ z_4,\bar{z}_4\right]=-\frac{\zeta}{2}\,,
\end{equation}
for a positive constant parameter $\zeta$.
These commutation relations can be represented using creation and annihilation 
operators\,;
\begin{eqnarray}
& &  \sqrt{\frac{2}{\zeta}} z_1\,|n_1:n_2:n_3:n_4\rangle =\sqrt{ n_1+1}\,
|n_1+1:n_2:n_3:n_4\rangle\,,\nonumber\\ 
& &  \sqrt{\frac{2}{\zeta}} \bar{z}_1\,|n_1:n_2:n_3:n_4\rangle =\sqrt{ n_1}\,
|n_1-1:n_2:n_3:n_4\rangle\,,\nonumber\\
& & \sqrt{\frac{2}{\zeta}} \bar{z}_2\,|n_1:n_2:n_3:n_4\rangle =\sqrt{ n_2+1}\,
|n_1:n_2+1:n_3:n_4\rangle\,,\nonumber\\
& &  \sqrt{\frac{2}{\zeta}} z_2\,|n_1:n_2:n_3:n_4\rangle =\sqrt{ n_2}\,
|n_1:n_2-1:n_3:n_4\rangle\,,\\
& & \sqrt{\frac{2}{\zeta}} z_3\,|n_1:n_2:n_3:n_4\rangle =\sqrt{ n_3+1}\,
|n_1:n_2:n_3+1:n_4\rangle\,,\nonumber\\ 
& & \sqrt{\frac{2}{\zeta}} \bar{z}_3\,|n_1:n_2:n_3:n_4\rangle =\sqrt{ n_3}\,
|n_1:n_2:n_3-1:n_4\rangle\,,\nonumber\\
& & \sqrt{\frac{2}{\zeta}} \bar{z}_4\,|n_1:n_2:n_3:n_4\rangle =\sqrt{ n_4+1}\,
|n_1:n_2:n_3:n_4+1\rangle\,,\nonumber\\ 
& & \sqrt{\frac{2}{\zeta}} z_4\,|n_1:n_2:n_3:n_4\rangle =\sqrt{ n_4}\,
|n_1:n_2:n_3:n_4-1\rangle\,.\nonumber
\end{eqnarray}

If we also require that $D^{\dagger}_zD_z$ commutes with $\Sigma_{\mu}$ 
as in the commutative case, 
we can obtain the ``self-dual'' gauge field strength on the 
noncommutative $\mathbf{R}^8$. 
As in the four dimensional case, there are also equivalence relations 
between different sets of matrices $\mathbf{A}\,,\,\mathbf{B}$ and 
$\mathbf{B}'$ .
Using these relations, the eight dimensional noncommutative ADHM equations 
 were proposed in~\cite{ohta}.
However it is even difficult to find simple explicit solutions 
of those equations.
Then instead of solving them, we solve the conditions for ``self-duality'' 
as~\cite{pt}, and give some simple solutions below.


\begin{flushleft}
\textbf{$U(1)$ one-instanton solutions}
\end{flushleft}

In the noncommutative case, $U(1)$ instanton is already non-trivial.
In this case, $D_z$ becomes a $3\times 2$ matrix.
If we consider the following ansatz\,;
\begin{equation}
D_z = \left( \begin{array}{@{\,}cc@{\,}}
  z_2 +z_4 &  z_1 +z_3\\
   -\bar{z}_1  -\bar{z}_3 & \bar{z}_2+\bar{z}_4\\
  A_1 & A_2
  \end{array}  \right)\,,
\end{equation}
the commuting condition of $D^{\dagger}_zD_z$ becomes
\begin{equation}
\left[ z_1,\bar{z}_1\right]-\left[ z_2,\bar{z}_2\right]
+ \left[ z_3,\bar{z}_3\right]-\left[ z_4,\bar{z}_4\right]
+A_1^{\dagger}A_1 -A_2^{\dagger}A_2=0\,,\quad  A_1^{\dagger}A_2=0\,.
\end{equation}
When $\zeta >0$, these equations have a non-trivial solution as
\begin{equation}
  A_1=\sqrt{2\zeta}\,,\quad A_2=0\,.
\end{equation}
Then $D_{z}$ becomes
\begin{equation}
D_z =
 \left( \begin{array}{@{\,}cc@{\,}}
  z_2 +z_4 &  z_1 +z_3\\
   -\bar{z}_1  -\bar{z}_3 & \bar{z}_2+\bar{z}_4\\
  \sqrt{2\zeta} & 0
  \end{array}  \right)\,.\label{eq:4.21}
\end{equation}
This is the simplest extension of the four dimensional $U(1)$ 
one-instanton solution found by~\cite{ns, furuuchi2}. 
Zero mode $\psi$ of the $D^{\dagger}_z$ is given by
\begin{equation}
\psi =  \left( \begin{array}{@{\,}c@{\,}}
   -\sqrt{2\zeta}(z_2+z_4)  \\
  \sqrt{2\zeta}(\bar{z}_1+\bar{z}_3)   \\
  (z_1+z_3)(\bar{z}_1+\bar{z}_3) +(\bar{z}_2+\bar{z}_4)(z_2+z_4)
  \end{array}  \right)\,.
\end{equation}
As in the four dimensional case, $\psi^{\dagger}\psi$ annihilates 
$|0:0:0:0\rangle\langle 0:0:0:0|$ 
so that we must normalize the zero mode in the subspace of the Fock space 
where $ |0:0:0:0\rangle$ is projected out~\cite{furuuchi, furuuchi2}.

We can generalize the above solution assuming the ansatz such as~\cite{pt}\,;
\begin{equation}
D_z =
\left( \begin{array}{@{\,}c@{\,}}
 (\mathbf{p})^{\dagger}\mathbf{x} 
 +(\mathbf{p}^{\prime})^{\dagger}
\mathbf{x}^{\prime} -\mathbf{a} \\
 A 
  \end{array}  \right)\,.
\end{equation}
Here $\mathbf{p}$, $\mathbf{p}^{\prime}$ and $\mathbf{a}$ are assumed 
to be arbitrary quaternions, and $A$ is a $1\times 2$ matrix.
Then the commuting condition of $D^{\dagger}_zD_z$  
can be solved when $\zeta >0$ as
\begin{equation}
A=\left( \begin{array}{@{\,}cc@{\,}}
\sqrt{\zeta \left( (\mathbf{p})^{\dagger}\mathbf{p} 
+(\mathbf{p}^{\prime})^{\dagger}\mathbf{p}^{\prime}\right) } & 0
  \end{array}  \right)\,.
\end{equation}
When $\mathbf{p}=\mathbf{p}^{\prime}=\mathbf{1}_{2}$ and 
$\mathbf{a}=\mathbf{0}_2$ the above solution 
reduces to (\ref{eq:4.21}), 
and when $\mathbf{p}^{\prime}=\mathbf{a}=\mathbf{0}$ 
and $\mathbf{p}=\mathbf{1}_{2}$ to the four dimensional noncommutative 
$U(1)$ one-instanton solution found by~\cite{furuuchi, furuuchi2} 
corresponding to $D4$-$D8$ bound states in this case.
If instead $\mathbf{p}=\mathbf{a}=\mathbf{0}$ 
and $\mathbf{p}^{\prime}=\mathbf{1}_{2}$, the solution reduces 
to the other four dimensional noncommutative 
$U(1)$ one-instanton solution.
Therefore our solutions are natural to extend the four dimensional solutions.


\begin{flushleft}
\textbf{$U(1)$ two-instanton solutions}
\end{flushleft}

In this case $D_z$ becomes a $5\times 4$ matrix. 
We assume the ansatz such as~\cite{pt}\,;
\begin{equation}
D_z =
 \left( \begin{array}{@{\,}cc@{\,}}
 (\mathbf{p}_1)^{\dagger}\mathbf{x}^1 +(\mathbf{p}_1^{\prime})^{\dagger}
\mathbf{x}^2 -\mathbf{a}_1 & 0 \\
 0 & (\mathbf{p}_2)^{\dagger}\mathbf{x}^1 
+(\mathbf{p}_2^{\prime})^{\dagger}\mathbf{x}^2 -\mathbf{a}_2 \\
 -\lambda_1 A_{1}  &  -\lambda_2 A_{2} 
  \end{array}  \right)\,,
\end{equation}
where $\mathbf{p}_{1,2}$, $\mathbf{p}^{\prime}_{1,2}$ 
and $\mathbf{a}_{1,2}$ are assumed 
to be arbitrary quaternions, $A_{1,\,2}$ are $1\times 2$ matrices, and 
$\lambda_{1,2}$ are constants.
Then if we take $\lambda_{1,2}$ as
\begin{equation}
\lambda _1 = \sqrt{ \zeta \left( (\mathbf{p}_1)^{\dagger}\mathbf{p}_1+
 (\mathbf{p}_1^{\prime})^{\dagger}\mathbf{p}_1^{\prime} \right)}\,,\quad
\lambda _2 = \sqrt{ \zeta \left( (\mathbf{p}_2)^{\dagger}\mathbf{p}_2+
 (\mathbf{p}_2^{\prime})^{\dagger}\mathbf{p}_2^{\prime}   \right)}\,,
\end{equation}
the commuting condition of $D_z^{\dagger}D_z$ becomes the following 
equations\,;
\begin{eqnarray}
& &  (A_{1}^{\dagger}A_{1})_{11}-
 (A_{1}^{\dagger}A_{1})_{22} =  1\,,\quad
  (A_{2}^{\dagger}A_{2})_{11}-
 (A_{2}^{\dagger}A_{2})_{22}  = 1 \,, \\
& & A_{1}^{\dagger}A_{2}\propto 1_{2\times 2}\,,
\quad  A_{2}^{\dagger}A_{1}\propto 1_{2\times 2}\,,\quad 
(A_{1}^{\dagger}A_{1})_{12}=(A_{2}^{\dagger}A_{2})_{12}=0\,.
\end{eqnarray}
Now $A_{1}$ and $A_{2}$ are $1\times 2$ matrices, 
these equations seem to have no non-trivial solutions.
However if we set $\mathbf{p}_1,\,\mathbf{p}_2$ or
$\mathbf{p}_1^{\prime},\,\mathbf{p}_2^{\prime}$ to zero, 
these equations reduce to the four dimensional $U(1)$ two-instanton 
ADHM equations, then can have non-trivial solutions. 
As we will see below, in the $U(2)$ case we can construct non-trivial 
eight dimensional two-instanton solutions using the above ansatz.


\begin{flushleft}
\textbf{$U(2)$ one-instanton solutions}
\end{flushleft}

In this case, $D_z$ becomes a $4\times 4$ matrix. 
Here also we assume the ansatz such as
\begin{equation}
D_z =
 \left( \begin{array}{@{\,}c@{\,}}
 (\mathbf{p})^{\dagger}\mathbf{x}^1 +(\mathbf{p}^{\prime})^{\dagger}
\mathbf{x}^2
 -\mathbf{a}\\
   -\lambda A   
  \end{array}  \right)\,,
\end{equation}
where $\mathbf{p}$, $\mathbf{p}^{\prime}$ 
and $\mathbf{a}$ are assumed 
to be arbitrary quaternions, 
 $A$ is a $2\times 2$ matrix which is not necessary to 
be quaternion, and $\lambda$ is a constant.
Then the commuting condition of $D_z^{\dagger}D_z$ becomes the following 
equations\,;
\begin{equation}
\lambda^2 \left( ( A^{\dagger}A)_{11}-(A^{\dagger}A)_{22}\right)= 
\zeta \left( (\mathbf{p})^{\dagger}\mathbf{p} + (\mathbf{p}^{\prime})^{\dagger}
\mathbf{p}^{\prime} \right)\,,\quad ( A^{\dagger}A)_{12}=0\,.
\end{equation}
If we take $\lambda$ as
\begin{equation}
\lambda =\sqrt{ \zeta \left( (\mathbf{p})^{\dagger}\mathbf{p} 
+(\mathbf{p}^{\prime})^{\dagger}\mathbf{p}^{\prime} \right) }\,,
\end{equation}
then a matrix $A$ is for example given by
\begin{equation}
A=
\left( \begin{array}{@{\,}cc@{\,}}
 0   &    0     \\
  1    &   0 
  \end{array}  \right)\,.
\end{equation}
This gives an eight dimensional noncommutative 
$U(2)$ one-instanton solution.


\begin{flushleft}
\textbf{$U(2)$ two-instanton solutions}
\end{flushleft}

In this case, $D_z$ is a $6\times 4$ matrix.
Here also we assume the ansatz such as
\begin{equation}
D_z =
\left( \begin{array}{@{\,}cc@{\,}}
 (\mathbf{p}_1)^{\dagger}\mathbf{x}^1 +(\mathbf{p}^{\prime}_1 )^{\dagger}
\mathbf{x}^2 -(\mathbf{a}_1) & 0 \\
0 & (\mathbf{p}_2 )^{\dagger} \mathbf{x}^1 +(\mathbf{p}^{\prime}_2) ^{\dagger}
\mathbf{x}^2 -(\mathbf{a}_2) \\
   -\lambda_1 A_{1}  &  -\lambda_2 A_{2}  
  \end{array}  \right)\,,\label{eq:4.33}
\end{equation}
where $\mathbf{p}_{1,2}$, $\mathbf{p}^{\prime}_{1,2}$ 
and $\mathbf{a}_{1,2}$ are assumed 
to be arbitrary quaternions, $A_{1,\,2}$ are $2\times 2$ matrices 
which are not necessary to 
be quaternions, and $\lambda_{1,2}$ are constants.
Then if we take $\lambda_{1,2}$ as 
\begin{eqnarray}
\lambda _1 &=& \sqrt{ \zeta \left(   
(\mathbf{p}_{1})^{\dagger}\mathbf{p}_{1} + (\mathbf{p}_{1}^{\prime})^{\dagger}
\mathbf{p}_{1}^{\prime} \right)}\,,\\
\lambda _2 &=& \sqrt{ \zeta \left( 
(\mathbf{p}_{2})^{\dagger}\mathbf{p}_{2} + (\mathbf{p}_{2}^{\prime})^{\dagger}
\mathbf{p}_{2}^{\prime} \right)}\,,
\end{eqnarray}
the commuting condition of $D_z^{\dagger}D_z$ becomes the following 
equations\,;
\begin{eqnarray}
& &  (A_{1}^{\dagger}A_{1})_{11}-(A_{1}^{\dagger}A_{1})_{22} = 1 \,,
\quad (A_{2}^{\dagger}A_{2})_{11}-(A_{2}^{\dagger}A_{2})_{22}  = 1 \,, \\
& & A_{1}^{\dagger}A_{2}\propto 1_{2\times 2}\,,\quad 
 A_{2}^{\dagger}A_{1}\propto 1_{2\times 2}\,,\quad 
(A_{1}^{\dagger}A_{1})_{12}=(A_{2}^{\dagger}A_{2})_{12}=0\,.
\end{eqnarray}
In contrast with the $U(1)$ two-instanton case, 
these equations have for example a non-trivial solution\,;
\begin{equation}
A_{1} = \left( \begin{array}{@{\,}cc@{\,}}
   0   &    0     \\
   1   &    0 
  \end{array}  \right)\,,\quad  A_{2} =
\left( \begin{array}{@{\,}cc@{\,}}
   1   &    0     \\
   0   &    0 
  \end{array}  \right)\,.
\end{equation}
 This gives an eight dimensional noncommutative 
$U(2)$ two-instanton solution.


\begin{flushleft}
\textbf{$U(2)$ $k$-instanton solutions}
\end{flushleft}

The above case can be easily generalized to multi-instanton solution. 
In this case, $D_{z}$ becomes a $(2+2k)\times (2k)$ matrix.
We consider the following ansatz\,;
\begin{equation}
D_z =
 \left( \begin{array}{@{\,}ccc@{\,}}
 (\mathbf{p}_1)^{\dagger}\mathbf{x}^1 +(\mathbf{p}_1^{\prime})^{\dagger}
\mathbf{x}^2 -(\mathbf{a}_1)
  &\ldots &  0  \\
   \vdots & \ddots  & \vdots  \\
   0  &\ldots &  
(\mathbf{p}_k)^{\dagger} \mathbf{x}^1 +(\mathbf{p}_k^{\prime})^{\dagger}
\mathbf{x}^2 -(\mathbf{a}_k)  \\
 -\lambda _1 A_{1} & \ldots & -\lambda_k A_{k} 
  \end{array}  \right)\,,
\end{equation}
where $\mathbf{p}_{1,\cdots,\,k}$, $\mathbf{p}^{\prime}_{1,\cdots,\,k}$ 
and $\mathbf{a}_{1,\cdots,\,k}$ are assumed 
to be arbitrary quaternions, $A_{1,\cdots,\,k}$ are $2\times 2$ matrices 
which are not necessary to 
be quaternions, and $\lambda_{1,\cdots,\,k}$ are constants.
Then if we take $\lambda_{1,\cdots,\,k}$ as 
\begin{eqnarray}
\lambda _1 &=& \sqrt{ \zeta \left(   
(\mathbf{p}_{1})^{\dagger}\mathbf{p}_{1} + (\mathbf{p}_{1}^{\prime})^{\dagger}
\mathbf{p}_{1}^{\prime} \right)}\,,\\
& & \vdots \nonumber\\
\lambda _k &=& \sqrt{ \zeta \left( 
(\mathbf{p}_{k})^{\dagger}\mathbf{p}_{k} + (\mathbf{p}_{k}^{\prime})^{\dagger}
\mathbf{p}_{k}^{\prime} \right)}\,,
\end{eqnarray}
the commuting condition of $D_z^{\dagger}D_z$ becomes the following 
equations\,;
\begin{eqnarray}
& &  (A_{1}^{\dagger}A_{1})_{11}-(A_{1}^{\dagger}A_{1})_{22} = 1 \,,
\cdots,\, (A_{k}^{\dagger}A_{k})_{11}-(A_{k}^{\dagger}A_{k})_{22}  = 1 \,, \\
& & A_{1}^{\dagger}A_{k}\propto 1_{2\times 2}\,,\quad 
 A_{2}^{\dagger}A_{k}\propto 1_{2\times 2}\,,\cdots\,,A_{k-1}^{\dagger}A_{k}
\propto 1_{2\times 2}\,,\\
& & (A_{1}^{\dagger}A_{1})_{12}=\cdots =(A_{k}^{\dagger}A_{k})_{12}=0\,.
\end{eqnarray}
These equations have for example a non-trivial solution\,;
\begin{equation}
A_{1} =A_{2}=\cdots =A_{k-1}= \left( \begin{array}{@{\,}cc@{\,}}
   0   &    0     \\
   1   &    0 
  \end{array}  \right)\,,\quad  A_{k} =
\left( \begin{array}{@{\,}cc@{\,}}
   1   &    0     \\
   0   &    0 
  \end{array}  \right)\,.
\end{equation}
 This gives an eight dimensional noncommutative 
$U(2)$ $k$-instanton solution.


\section{Discussions}

In this paper, 
we described eight dimensional noncommutative 
Yang-Mills instantons associated with the group $Sp(2)$ 
as $D0$-$D8$ bound states in a $B$-field
using the noncommutative eight dimensional ADHM construction. 
In this case, the $B$-field was necessary to satisfy the extended 
``(anti-)self-dual'' relations for preserving some proportions of supercharges.
We constructed some new classes of non-trivial solutions 
of the eight dimensional noncommutative ADHM equations 
generalizing the solutions found by Papadopoulos and Teschendorff~\cite{pt} 
in the commutative case, 
and interpreted them as $D0$-$D8$ bound states.

However in the noncommutative case 
it is non-trivial whether the instanton number 
is actually the expected one, therefore it would be interesting to calculate  
explicitly the instanton numbers.

In the commutative case, (\ref{eq:4.33}) is a solution of eight dimensional 
ADHM equations when $A_{1}(=A_{2})$ is 
arbitrary quaternion and $\lambda_{1,\,2}$ 
are arbitrary constants. Then there exists following equivalence relations\,;
\[  \{ \mathbf{p}_{1},\,\mathbf{p}_1^{\prime},\,\mathbf{p}_{2},\,
\mathbf{p}_2^{\prime},\,\mathbf{a}_1^{\dagger},\,\mathbf{a}_2^{\dagger} ,\,
A_1^{\dagger}\}
\sim  \{ \mathbf{p}_{1}\mathbf{s},\,\mathbf{p}_1^{\prime}\mathbf{s},\,
\mathbf{p}_{2}\mathbf{s},\,
\mathbf{p}_2^{\prime}\mathbf{s},\,\mathbf{a}_1^{\dagger}\mathbf{s},\,
\mathbf{a}_2^{\dagger}\mathbf{s},\,A_1^{\dagger}\mathbf{s} \}\,,\]
where $\mathbf{s}\in \mathbf{H}$. 
Therefore the moduli space structure of the above solution  
may be related with the Grassmannian\,;
\[  \frac{Sp(2)}{Sp(1)\times Sp(1)} \,, \] 
and this may suggest that the moduli space of the eight dimensional 
instanton will have an $Sp(2)$ holonomy~\cite{ohta}. 
It is not yet made clear 
how the noncommutativity deforms the moduli space structure of 
the eight dimensional instantons, but 
we expect that our solutions will shed light on this problem. 

It is also of interest to search for the ADHM constructions
associated with the group $Spin(7)$ and $SU(4)$.
In the $Spin(7)$ case, some solutions are known 
as octonionic Yang-Mills instantons in~\cite{ky} which are 
constructed on the $Spin(7)$ holonomy manifold (gravitational instanton) 
by the standard embedding of the spin connection into 
the Yang-Mills connection.
In four dimensional case, the ADHM constructions of instantons 
on an ALE space (gravitational instanton) are well-known~\cite{dm}.
Eight dimensional ALE spaces are considered in~\cite{joyce, joyce2}.
Therefore we expect that similar constructions are possible 
in the eight-dimensional case, and our solutions give some insight into 
the moduli space structures of eight dimensional instantons.


\bigskip
\bigskip
\centerline{\bf Acknowledgments}

\vskip 0.6cm

We would like to thank K.Ohta for useful comments 
and H.Ishikawa for useful discussions and 
encouragements.


\noindent
\bibliography{}

\begin{thebibliography}{999}

\bibitem{cds}
A.Connes, M.R.Douglas and A.Schwarz, ``Noncommutative Geometry and Matrix
 Theory: Compactification on Tori'', \textbf{JHEP 9802} (1998) 003, 
hep-th/9711162.

\bibitem{sw}
N.Seiberg and E.Witten, ``String Theory and the Noncommutative Geometry'',
 \textbf{JHEP 9909} (1999) 032, hep-th/9908142.

\bibitem{ns}
N.Nekrasov and A.Schwarz, ``Instantons on noncommutative $\mathbf{R}^4$, and
 (2,0) superconformal six dimensional theory'', Comm. Math. Phys. \textbf{198}
 (1998) 689-703, hep-th/9802068.

\bibitem{nekrasov}
N.Nekrasov, ``Trieste lectures on solitons in noncommutative gauge theories'',
 hep-th/0011095.

\bibitem{furuuchi}
K.Furuuchi, ``Topological Charges of $U(1)$ Instantons'', hep-th/0010006.

\bibitem{furuuchi2}
K.Furuuchi, ``Instantons on Noncommutative $\mathbf{R}^4$ and Projection 
Operators'', Prog. Theor. Phys. \textbf{103} (2000) 1043-1068, hep-th/9912047.

\bibitem{furuuchi3}
K.Furuuchi, ``Dp-D(p+4) in Noncommutative Yang-Mills'', 
JHEP \textbf{0103} (2001) 033, hep-th/0010119.

\bibitem{cimm}
B.Chen, H.Itoyama, T.Matsuo and K.Murakami, ``$p$-$p'$ System with $B$-field,
 Branes at angles and Noncommutative Geometry'', Nucl. Phys. \textbf{B576}
 (2000) 177-195, hep-th/9910263.

\bibitem{park}
M.Mihailescu, I.Y.Park and T.A.Tran, 
``D-branes as Solitons of an N=1, D=10 Non-commutative Gauge Theory'', 
Phys. Rev. \textbf{D64} (2001) 046006, hep-th/0011079.

\bibitem{witten}
E.Witten, ``BPS Bound States Of $D0$-$D6$ And $D0$-$D8$ Systems 
In A $B$-field'',
 hep-th/0012054.

\bibitem{fio}
A.Fujii, Y.Imaizumi and N.Ohta, ``Supersymmetry, Spectrum and Fate of D0-D$p$ 
Systems with $B$-field'', Nucl. Phys. \textbf{B615} (2001) 61-81, 
hep-th/0105079.

\bibitem{cdfn}
E.Corrigan, C.Devchand, D.B.Fairlie and J.Nuyts, ``First-order Equations for
 Gauge Fields in Space of Dimension Greater than Four'', Nucl. Phys.
 \textbf{B214} (1983) 452-464.

\bibitem{ward}
R.S.Ward, ``Completely Solvable Gauge-field Equations In Dimension
Greater Than Four'', Nucl. Phys. \textbf{B239} (1984) 381-396.

\bibitem{hull}
C.M.Hull, ``Higher Dimensional Yang-Mills Theories and Topological Terms'', 
Adv. Theor. Math. Phys. \textbf{2} (1998) 619-632 hep-th/9710165.

\bibitem{adhm}
M.Atiyah, N.Hitchin, V.Drinfeld and Y.Manin, ``Construction of Instantons'',
 Phys. Lett. \textbf{65B} (1978) 185.

\bibitem{cg}
E.Corrigan and P.Goddard, ``Construction of Instanton and Monopole Solutions
 and Reciprocity'', Ann.Phys. \textbf{154} (1984) 253-279.

\bibitem{cgk}
E.Corrigan, P.Goddard and A.Kent, ``Some Comments on the ADHM Construction
 in $4k$ Dimensions'', Comm. Math. Phys. \textbf{100} (1985) 1-13.

\bibitem{ohtan}
N.Ohta and P.K.Townsend, ``Supersymmetry of M-Branes at Angles'', 
Phys. Lett. \textbf{B418} (1998) 77-84, hep-th/9710129.

\bibitem{ohta}
K.Ohta, ``Supersymmetric D-brane Bound States with $B$-field and
Higher Dimensional Instantons on Noncommutative Geometry'',
 Phys. Rev. \textbf{D64} (2001) 046003,
 hep-th/0101082.

\bibitem{hio}
M.Hamanaka, Y.Imaizumi and N.Ohta, ``Moduli Space and Scattering of D0-branes 
in Noncommutative Super Yang-Mills Theory'', Phys. Lett. 
\textbf{B529} (2002) 163-170, hep-th/0112050.

\bibitem{pt}
G.Papadopoulos and A.Teschendorff, ``Instantons at Angles'', Phys. Lett.
 \textbf{B419} (1998) 115-122, hep-th/9708116.

\bibitem{pt2}
G.Papadopoulos and A.Teschendorff, ``Multi-angle Five-Brane Intersections'',
 Phys. Lett. \textbf{B443} (1998) 159-166, hep-th/9806191.

\bibitem{ggpt}
J.P.Gauntlett, G.W.Gibbons, G.Papadopoulos and P.K.Townsend, 
``Hyper-K\"{a}hler manifolds and multiply-intersecting branes'', 
Nucl. Phys. \textbf{B500} (1977) 133-162, hep-th/9702202.

\bibitem{ky}
H.Kanno and Y.Yasui, ``Octonionic Yang-Mills Instanton on Quaternionic Line 
Bundle of Spin(7) Holonomy'', J.Geom.Phys. \textbf{34} (2000) 302-320 
hep-th/9910003.

\bibitem{dm}
M.R.Douglas and G.Moore, ``D-branes, Quivers and ALE instantons'', 
hep-th/9603167.

C.I.Lazaroiu, ``A noncommutative-geometric interpretation of the 
resolution of equivalent instanton moduli spaces'', hep-th/9805132.

\bibitem{joyce}
D.Joyce, ``Asymptotically Locally Euclidean metrics with holonomy SU(m)'', 
math.AG/9905041.

\bibitem{joyce2}
D.Joyce, ``Quasi-ALE metrics with holonomy SU(m) and Sp(m)'', 
math.AG/9905043.


\end{thebibliography}

\end{document}